# HIGH LUMINOSITY INTERACTION REGION DESIGN FOR COLLISIONS WITH DETECTOR SOLENOID*

C. Milardi, M. Preger, P. Raimondi, G. Sensolini, F. Sgamma, LNF-INFN, Frascati, Italy.

*Abstract*

An innovatory interaction region has been recently conceived and realized on the Frascati DAΦNE lepton collider. The concept of tight focusing and small crossing angle adopted until now to achieve high luminosity in multibunch collisions has evolved towards enhanced beam focusing at the interaction point with large horizontal crossing angle, thanks to a new compensation mechanism for the beam-beam resonances. The novel configuration has been tested with a small detector without solenoidal field yielding a remarkable improvement in terms of peak as well as integrated luminosity. The high luminosity interaction region has now been modified to host a large detector with a strong solenoidal field which significantly perturbs the beam optics introducing new design challenges in terms of interaction region optics design, beam transverse coupling control and beam stay clear requirements.

## INTRODUCTION

Recently a new collision scheme based on large Piwinski angle and Crab-Waist compensation of the beam-beam interaction [1] has been proposed and implemented [2] on DAΦNE [3].

The new configuration has been used to provide beam-beam events to the SIDDHARTA [4] experiment, a compact device without solenoidal field, heir of DEAR, providing a simple environment for the Crab-Waist test. The luminosity has been increased by a factor 3 with a peak value of $4.53 \cdot 10^{32}$ cm$^{-2}$s$^{-1}$ letting in collision currents slightly lower than those corresponding to the old records. The highest daily integrated luminosity measured in a moderate injection regime, suitable for SIDDHARTA operation, has been $L_{\int day}$ ~15 pb$^{-1}$. An almost continuous injection regime provided $L_{\int 1hour}$ ~1.0 pb$^{-1}$ hourly integrated luminosity which opened significant perspectives for the KLOE-2 experiment. Scaling this best integrated luminosity measured over two hours, it is reasonable to expect more than 20 pb$^{-1}$ per day, and assuming 80% collider uptime as during the past runs, ~ 0.5 fb$^{-1}$ per month [5, 6].

The results of the high luminosity test have renovated the interest about the experimental activity on the DAΦNE collider, paving the way for a new run [7] with an upgraded KLOE detector, KLOE-2 [8].

*Work supported by the EuCARD research programme, within the "Assessment of Novel Accelerator Concepts" work package (ANAC-WP11).

## THE DAΦNE INTERACTION REGION FOR THE KLOE-2 RUN

Integrating the high luminosity collision scheme with the KLOE-2 detector introduces new challenges in terms of Interaction Region (IR) layout and optics, beam acceptance and coupling correction [9].

### IR Layout and Vacuum Chamber

The low-β section is based, as for the SIDDHARTA configuration, on permanent magnet quadrupole doublets. The quadrupoles are made of SmCo alloy and provide gradients of 29.2 T/m for the first one from the IP and 12.6 T/m for the second one.

The first (PMQD) is horizontally defocusing and is shared by the two beams; its central azimuthal position has been set at 0.415 m from the IP, as a compromise between the conflicting requirements of tight focusing and large solid angle aperture coming from the collider and the experiment respectively. The second quadrupole (PMQF), horizontally focusing, is installed just after the point where the beam pipes of the two rings are separated and is therefore on axis. Being PMQD much stronger than in the old KLOE low-β setup and having doubled the horizontal half crossing angle, now ~25 mrad, a very efficient beam separation is achieved in the ~1.6 m long section of the IR common to the two rings, making the impact of a single parassitic crossing completely negligible. As a drawback the horizontal and vertical displacement of the beam in the IR, the latter strongly affected by the detector solenoidal field, becomes an order of magnitude larger than in the past KLOE run. To keep the beam vertical trajectory within reasonable values a permanent magnet dipole, PMD, has been added just after PMQF, inside the detector magnetic field, in each one of the four IR branches. Each PMD is built by using a SmCo alloy, consists of two parts having 75.0 mm magnetic length each, and provides an integrated field strength BL = 0.0168 Tm corresponding to a vertical deflection angle of ~10.0 mrad. The PMDs are based on a modular design in view of a possible KLOE-2 run at a lower solenoidal field. They provide a horizontal magnetic field directed inward in the e$^+$ ring and outward in the e$^-$ one, as shown in Fig. 1.

The IR magnetic layout, sketched in Fig. 1, has been designed in order to maximize the beam stay clear letting the beam trajectory pass as much as possible through the center of the magnetic elements according to a self-consistent procedure which, for symmetry reasons, has been developed for one section of the IR only. The crossing angle has been tuned in order to have the same horizontal displacement as in the SIDDHARTA configuration at the corrector dipole, DHCPS01, used to match the IR to the ring layout in the arcs, under the constraint of placing the PMDs as close as possible to the PFQMs; its present half value is $|\theta_c| = 25.7$ mrad. The longitudinal coordinates of the magnetic elements have been set to optimize the IR optics, while their transverse position and tilt have been adjusted to match the nominal beam orbit at the entrance of anti-solenoid COMPS001.

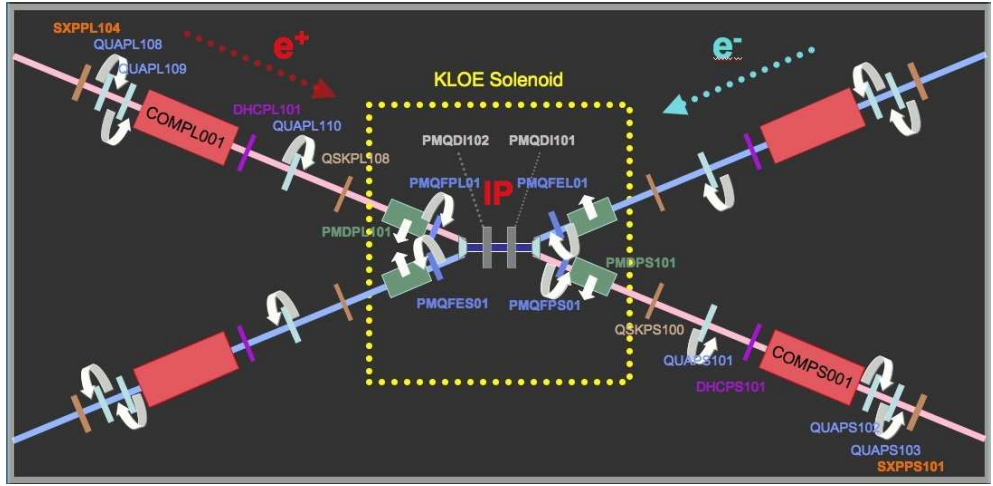

Figure 1: Schematic drawing of the DAFNE Interaction.

The beam trajectory and positions of the magnetic element centers are shown in Fig. 2 with reference to the IR branch of the positron ring pointing to the short arc, the corresponding branch for the electron ring being symmetric. The QUAPS101 quadrupole is slightly displaced with respect to the beam in order to compensate the small steering due to the practical constraint of leaving it in the upright position for alignment reason.

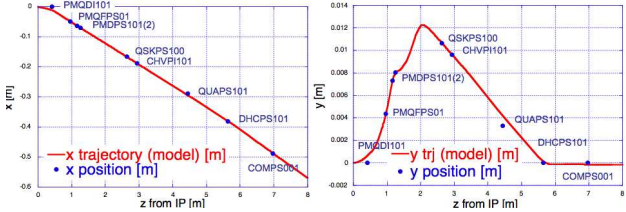

Figure 2: Horizontal (left) and Vertical (right) beam trajectory in the IR (solid line), dots represent the position of the magnetic element centers.

The evident advantage of this approach consists in keeping the maximum excursion of the vertical beam trajectory within ~12 mm providing, at the same time, the maximum aperture for the beam.

The horizontal and vertical beam stay-clear requirements have been defined as:

$$X_{SC} = x_{trj} \pm 10\sigma_x$$
$$Y_{SC} = y_{trj} \pm 10\sigma_y$$

where $\sigma_x$ and $\sigma_y$ are the horizontal and vertical rms beam sizes respectively. Their values, computed with the collider emittance $\varepsilon = 0.4 \cdot 10^{-6}$ m for the horizontal plane and full coupling for the vertical one, are presented in Fig. 3 together with the vacuum pipe profile evaluated with respect to the beam trajectory. The beam envelope shown represents an upper limit: in fact it has been obtained by using a quite large value for the beam emittance, the one that will be used to restart the collider operations. Relaying on this analysis the radius of the vacuum pipe, in the section between the IP and the DHC, has been reduced. It is now 2.75 cm, while it was 4.4 cm during the SIDDHARTA run. A narrower vacuum chamber contributes to lower the ring impedance budget, to minimize the strength of trapped High Order Modes and to shift their frequencies away from the beam spectral lines [10].

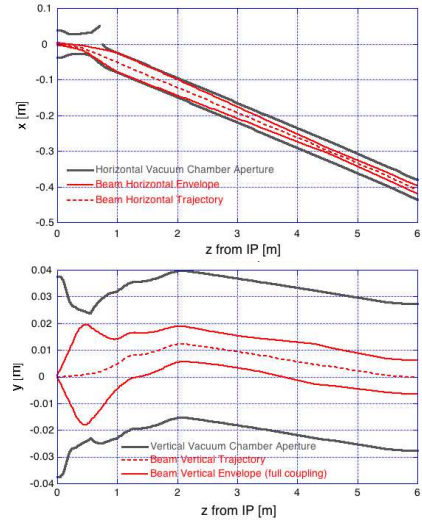

Figure 3: Horizontal (upper) and Vertical (lower) beam stay clear.

The resulting vacuum pipe geometry is largely simplified; in fact it consists of three straight sections, with few junctions and bellows. The detector efficiency also profits from the larger free space around the IP where a precision vertex tracker can be placed. The IR pipe is aluminum (AL6082) made with the exception of the sphere surrounding the IP, which is built in ALBEMET.

Such a structure could trap HOMs and for this reason it is shielded from to the beam by means of a Be cylinder. To minimise K meson regeneration the shield thickness has been almost halved (35 μ instead of the 65 μ of the last KLOE run).

Additional W screens have been added just before the PMQDs quadrupoles (w.r.t. the IP) to shield the experimental detector relying on the knowledge about the background source and distribution developed by means of measurements and studies undertaken during the SIDDHARTA run.

### IR Optics

The design of the IR optics is constrained by several criteria. It must provide the prescribed low-β parameters at the IP ($\beta_x = 0.265$ m, $\beta_y = 0.0085$ m, $\alpha_x = \alpha_y = 0.0$, $\eta_x = \eta'_x = 0.0$), matching at the same time the ring original layout in the arcs. The phase advance between the Crab-Waist Sextupoles and the IP must be π for the horizontal-like mode and π/2 for the vertical one. The value of the β-functions at the Crab-Waist Sextupoles must be tuned to the values that fully exploit the strength of the existing devices.

The transverse coupling introduced by the detector solenoid must be carefully compensated by means of the compensator solenoids and rotation of the IR quadrupoles.

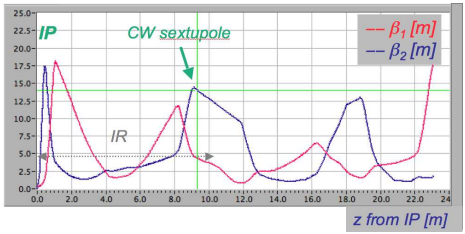

Figure 4: IR optical function.

### Coupling Correction

The transverse betatron coupling has been corrected revising the Rotating Frame Method [11] originally used at DAΦNE. The field integral introduced by the solenoidal detector is almost cancelled by means of two anti-solenoids, installed symmetrically with respect to the IP in each ring, which provide compensation also for off-energy particles. The rotation of the beam transverse plane is compensated by rotating the quadrupoles PMQFPS01, QUAPS101, QUAPS102, QUAPS103 around their longitudinal axis. The first low beta quadrupole has been kept in the upright position, since the nominal rotation suitable for coupling correction significantly increases the displacement of the beam vertical trajectory. The anti-solenoidal field has been set to a value slightly lower with respect to the optimal one in order to minimize the rotation angles and to make the tilts of the last two electromagnetic quadrupoles symmetric. It is worth remarking that the coupling is carefully compensated before the CW sextupoles making the terms of the coupling matrix vanish at QUAPS103.

Coupling correction fine-tuning, for each beam, is assured by a skew quadrupole added in each branch of the IR and by the two independently powered anti-solenoids.

The present coupling correction method introduces several advantages with respect to the one adopted for the past KLOE run. The rotation of the permanent magnet quadrupoles installed inside the detector is drastically reduced; in fact it was 8.3 deg and 12.9 deg for PMQD and PMQF respectively. Moreover it becomes possible to get rid of the remotely controlled actuators [12] used in the past to finely adjust the low-β quadrupole tilts, thus further increasing the solid angle available for the detection of decaying particles. Quadrupole rotations in the four sections of the IR are summarized in Tab. 2. They are defined so that a positive tilt gives a clockwise rotation in the reference system moving with the positron beam.

A residual coupling in the range $\kappa = 0.2\% \div 0.3\%$, as in the past KLOE run is expected to be reached by this approach. The IR coupling correction related parameters are summarized in Tab. 2.

Table 2: Coupling Correction parameters

| | Z from the IP [m] | Quadrupole rotation angles[deg] / Anti-solenoid current [A] |
|---|---|---|
| PMQDI101 | 0.415 | 0.0 |
| PMQFPS01 | 0.963 | -4.48 |
| QSKPS100 | 2.634 | used for fine tuning |
| QUAPS101 | 4.438 | -13.73 |
| QUAPS102 | 8.219 | 0.906 |
| QUAPS103 | 8.981 | -0.906 |
| COMPS001 | 6.963 | 72.48 (optimal value 86.7) |

## CONCLUSIONS

The high luminosity IR for the KLOE-2 detector has been designed, built and installed on the DAΦNE collider. All the different aspects related to layout, beam acceptance, optics and coupling correction have been studied in detail and optimized.

Transverse coupling correction is achieved by a simple scheme independently tunable for the two beams.

The IR optics has been defined in order to fulfill all the requirements in terms of low beta functions at the IP, coupling correction and inclusion of the crab sextupoles. Tracking studies have been performed to ensure compatibility of the new IR structure with the mechanical layout of the ring arcs.

The IR vacuum chamber mechanical design meets the requirements of lower impedance and larger free solid angle set by the collider and the experimental detector respectively.

## REFERENCES

[1] P.Raimondi, D.Shatilov, M.Zobov, physics/0702033.
[2] C.Milardi et al., arXiv:0803.1450v1 [physics.acc-ph].
[3] G. Vignola et al., Frascati Phys. Ser. 4:19-30,1996.
[4] SIDDHARTA Coll., Eur.J.Phys. A31:537-539,2007.
[5] C.Milardi et al., PAC09, Vancouver, MO4RAI01.
[6] M.Zobov et al., Phys.Rev.Lett.104:174801, 2010.
[7] C.Milardi et al., TUPEB006, this conference.
[8] F.Bossi, Journal of Physics: 171 (2009) 012099.
[9] C.Milardi et al., DAΦNE Technical Notes, IR-14.
[10] F.Marcellini, PAC07, 3988.
[11] M.Bassetti et al., Frascati Phys. Ser. Vol. X, 209, 98.
[12] C.Milardi et al., EPAC04, 233-235.